\documentclass[11pt,journal]{IEEEtran}

\usepackage{mathrsfs}
\usepackage{amsmath}
\usepackage{amssymb}
\usepackage{microtype}
\usepackage{graphicx}
\usepackage{subfigure}
\usepackage{booktabs} 
\usepackage{cite}
\usepackage[numbers]{natbib}

\usepackage{hyperref}
\usepackage{breakurl}

\def\gramloss{\mathcal{L}_{\textrm{{\scriptsize Gram}}}}
\def\autocorrelationloss{\mathcal{L}_{\textrm{{\scriptsize autocorr}}}}
\def\diversityloss{\mathcal{L}_{\textrm{{\scriptsize div}}}}
\def\sendikloss{\mathcal{L}_{\textrm{{\scriptsize Sendik}}}}
\def\vggishscore{\mathcal{S}_{\textrm{{\scriptsize VGGish}}}}

\def\citet{\cite}

\title{Synthesizing Diverse, High-Quality Audio Textures}
\author{
  Joe~Antognini\thanks{Work done as a Google AI Resident},
  Matt~Hoffman,
  Ron~J.~Weiss
}

\pubid{0000--0000/00\$00.00˜\copyright˜2018 IEEE}

\graphicspath{figs}

\begin{document}

\maketitle

\begin{abstract}

  Texture synthesis techniques based on matching the Gram matrix of feature
  activations in neural networks have achieved spectacular success in the
  image domain.  In this paper we extend these techniques to the audio
  domain.  We demonstrate that synthesizing diverse audio textures is
  challenging, and argue that this is because audio data is relatively
  low-dimensional.  We therefore introduce two new terms to the original
  Grammian loss: an autocorrelation term that preserves rhythm, and a
  diversity term that encourages the optimization procedure to synthesize
  unique textures.  We quantitatively study the impact of our design choices
  on the quality of the synthesized audio by introducing an audio analogue
  to the Inception loss which we term the VGGish loss.  We show that there
  is a trade-off between the diversity and quality of the synthesized audio
  using this technique.  We additionally perform a number of experiments to
  qualitatively study how these design choices impact the quality of the
  synthesized audio.  Finally we describe the implications of these results
  for the problem of audio style transfer.

\end{abstract}


\section{Introduction}
\label{sec:introduction}

Texture synthesis has been studied for over fifty years \cite{julesz62}.
The problem is to take a sample of some textured data (usually an image) and
generate synthesized data which have the same texture, but are not identical
to the original sample.  This problem is interesting as a machine learning
problem in its own right, but successful texture synthesis methods can also
elucidate the way in which humans perceive texture
\cite{mcdermott+simoncelli11}.

Portilla and Simoncelli \cite{portilla+simoncelli00} pioneered a very
successful approach to image texture synthesis that tries to find a complete
set of statistics that describe the perceptually relevant aspects of a given
texture.  To synthesize a new texture, a random input is perturbed until its
statistics match those of the targets.  Portilla and Simoncelli
\citet{portilla+simoncelli00} developed a set of four classes of statistics
consisting of 710 parameters which produced extremely realistic images of
natural and synthetic textures.  McDermott and Simoncelli
\citet{mcdermott+simoncelli11} used a similar approach to develop four
classes of statistics in a cochlear model to synthesize textural audio
spectrograms.  This work produced convincing audio data for many natural
audio textures (e.g., insects in a swamp, a stream, applause), but had
difficulty with pitched and rhythmic textures (e.g., wind chimes, walking on
gravel, church bells).

Gatys et al.~\citet{gatys+15a} introduced an extremely successful technique
that replaced hand-crafted statistics with Gram-matrix statistics derived
from the hidden feature activations of a trained convolutional neural
network (CNN).  By perturbing a random input to match these Gram matrices,
Gatys et al.~\citet{gatys+15a} produced compelling textures that were far
more complicated than those achieved by any earlier work.

Given the success of Gatys et al.~\citet{gatys+15a} in the image domain
relative to the hand-crafted approach of Portilla and Simoncelli
\citet{portilla+simoncelli00}, it is natural to ask whether a similar
CNN-based strategy could be adapted to the audio domain to build on the
hand-crafted approach of McDermott and Simoncelli
\citet{mcdermott+simoncelli11}. Ulyanov and Lebedev
\citet{ulyanov+lebedev16} proposed just such an extension of the approach of
\citet{gatys+15a} to audio. Their basic approach works fairly well on many
of the 15 examples they consider, but (as we demonstrate in
Sec.~\ref{sec:experiments}) it has some of the same failure modes as the
approach of McDermott and Simoncelli \citet{mcdermott+simoncelli11}.

In this work, we examine the causes of these problems, analyze why
they are more serious in the audio domain than in the image domain,
and propose techniques to fix them.



\pubidadjcol

\section{Preliminaries and Analysis}
\label{sec:preliminaries}

We begin by defining a rigorous notion of ``audio texture''.  Borrowing from
Portilla and Simoncelli \citet{portilla+simoncelli00}, we define an audio
texture to be an infinitely long stationary random process that follows an
exponential-family (maximum-entropy) distribution defined by a set of local
sufficient-statistic functions $\phi$ that are computed on patches of size
$M$:
\begin{equation}
  \label{eq:julesz}
  \begin{split}
    \textstyle
    p_\lambda(x)\propto\exp\{\lambda^\top \sum_{t=0}^\infty \phi(x_{t,\ldots,t+M})\},
    \; \mathbb{E}_{p_\lambda}[\phi] = \bar{\phi}.
\end{split}
\end{equation}
We only ever observe finite clips from this
theoretically infinite signal. If an observed clip is long enough
relative to $M$ and the dimensionality of $\phi$, then we can reliably
estimate the expected sufficient statistics $\bar{\phi}$ from data.

The first practical question we face is whether to model the data in the
time domain (i.e., the raw waveform) or the frequency domain (via a
spectrogram).  Although direct time-domain modeling has seen enormous
success in recent years with WaveNet \cite{van-den-oord+16}, these
autoregressive techniques are still slow to train and extremely
computationally demanding at synthesis-time.  Following Ulyanov and Lebedev
\citet{ulyanov+lebedev16}, we instead use the spectrogram representation in
this work.  This incurs the disadvantage of needing to invert the
spectrogram to recover audio with the Griffin-Lim algorithm
\cite{griffin+lim84}, which can introduce artifacts, but is much faster and
permits the unsupervised techniques we use in this work, eliminating the
need to train a large model.

Naively, an audio spectrogram can be treated like a two-dimensional greyscale image,
with time on one axis and frequency on the other. But in texture synthesis,
it is more natural to treat frequency bins in an audio spectrogram as channels
(analogous to RGB color channels in an image) rather than spatial dimensions.
Audio textures are not stationary on the frequency axis---shifts in frequency
tend to change the semantic meaning of a sound.
Therefore, although the ``spectrogram-as-image'' interpretation
can work well for some analysis problems \cite{hershey+17},
it is a poor fit to the assumptions underlying texture synthesis.

Treating spectrograms as one-dimensional multi-channel stationary signals
raises an important statistical issue that is not as salient in image
texture synthesis.  Images typically have only three channels, whereas audio
spectrograms have as many channels as there are samples in the FFT window
divided by two (typically some power of two between 128 and 2048).
Furthermore, the number of patches that are averaged to estimate the
statistics that define the target texture distribution is on the order of
the length of the signal, whereas in images the number of patches grows as
the product of the dimensions. So in audio texture synthesis, we need to
estimate a function of more channels with fewer observations, which may lead
to overfitting.  We indeed find that synthesizing diverse audio textures is
more difficult than synthesizing diverse images and extra care must be taken
to encourage diversity (with tongue planted in cheek, this may perhaps be
called a curse of low dimensionality).

\section{Methods}
\label{sec:methods}

\subsection{Signal processing}
\label{subsec:signal_processing}

We produce audio textures by transforming the target audio to a log
spectrogram and synthesizing a new spectrogram.  We then use the Griffin-Lim
algorithm \cite{griffin+lim84} to invert the spectrogram and generate the
synthesized audio texture.  If necessary, we resample the target audio to 16
kHz and normalize.  We produce a spectrogram by taking the absolute value of
the short-time Fourier transform with a Hann window of size 512 samples and
a hop size of 64 samples.  Although taking the absolute value removes any
explicit phase information, if the hop size is less than or equal to half
the window size phase information is implicitly retained (i.e., there exists
a unique audio signal corresponding to such a spectrogram up to a global
phase; \cite{sturmel+daudet11}).  We then add 1 to every magnitude in the
spectrogram and take the natural logarithm.  Adding 1 guarantees that the
log-spectrogram is finite and positive.

\subsection{Architecture of the neural networks}
\label{subsec:architecture}

We obtained the best textures with a set of six single-hidden-layer random
CNNs.  Unlike the case of image texture synthesis, audio spectrograms are
one-dimensional so we therefore use a one-dimensional convolution.  Each CNN
had a convolutional kernel with a different width, varying in powers of 2
from 2 to 64 frames.  We applied a ReLU activation after the convolutional
layers.  Each layer had 512 filters randomly drawn using the Glorot
initialization procedure \cite{glorot+bengio10}. Several authors have found
that random convolutional layers perform as well as trained convolutional
layers for image texture synthesis \cite{he+16, ustyuzhaninov+16}.  Shu et
al.~\citet{shu+17} furthermore showed that a random CNN retains as much
information to reconstruct an image as a trained convolutional network, if
not more.  Although we also tried synthesizing textures with an audio model
that was trained on AudioSet \cite{gemmeke+17}, a dataset consisting of
about one million 10 second audio clips with 527 labels, we did not find
that this trained model produced textures that were any better than those
produced by a random CNN.

Using an ensemble of CNNs with varying kernel sizes is crucial for obtaining
high quality textures since each kernel size is most sensitive to audio
features whose duration is comparable to the kernel size.  The features of
real-world audio can span many different timescales (e.g., just a few
milliseconds for a clap and up to several seconds for a bell) so it is
important to use an architecture which is sensitive to the range of
timescales that is likely to be encountered.  We consider the impact of our
architecture design choices experimentally in
Section~\ref{subsec:nn_architecture}.

\subsection{Loss terms}
\label{subsec:loss}

The loss we minimize consists of three terms:
\begin{equation}
  \label{eq:loss}
  \mathcal{L} = \gramloss +
  \alpha \autocorrelationloss +
  \beta \diversityloss.
\end{equation}

The first term, $\gramloss$, was introduced by Gatys et
al.~\citet{gatys+15a} and is intended to capture the average local
correlations between features in the texture.  The second term,
$\autocorrelationloss$, we adapt from Sendik and Cohen-Or
\citet{sendik+cohen-or17} and is intended to capture rhythm.  The final
term, $\diversityloss$, we introduce to prevent the optimization process
from exactly copying the original texture.  The hyperparameters $\alpha$ and
$\beta$ are used to set the relative importance of these three terms.  We
find that $\alpha = 10^{3}$ and $\beta = 10^{-4}$ work well for many of the
textures we studied, although hyperparameter tuning is sometimes required.
In particular, highly rhythmic textures generally require a larger choice of
$\alpha$ and a lower choice of $\beta$.

\subsubsection{Gram loss}
\label{subsubsec:gram_loss}

Let us write the features of the $k$th convolutional network as $F_{t \mu}^k$,
where $t$ indicates the position of a patch in the feature map (i.e., the
time in the spectrogram), and $\mu$ indicates the filter.  The Gram matrix for
the $k$th convolutional network is the time-averaged outer product between the $k$th
feature map with itself:
\begin{equation}
  \label{eq:avg_gram_matrix}
  G_{\mu \nu}^{k} = \frac{1}{T}\sum_t F_{t \mu}^{k} F_{t \nu}^{k},
\end{equation}
where $T$ is the number of windows in the spectrogram


We match this statistic by minimizing the Frobenius norm of the
difference between the Gram matrices of the synthesized texture and the
target for all layers and normalizing to the Frobenius norm of the target
texture Gram matrix:
\begin{equation}
  \textstyle
  \gramloss = \frac{\sum_{k, \mu, \nu} \left( G_{\mu \nu}^k -
  \widetilde{G}_{\mu \nu}^k \right)^2}
  {\sum_{k, \mu, \nu} \left( \widetilde{G}_{\mu \nu}^k \right)^2}.
\end{equation}
Throughout this paper tilde denotes the target texture.

\subsubsection{Autocorrelation loss}
\label{subsubsec:autocorrelation_loss}

While minimizing the Gram loss alone produces excellent audio for many kinds
of audio textures, we show in
Sec.~\ref{subsubsec:autocorrelation_experiments} that the Gram loss fails to
capture rhythm.  To this end, we adapt a loss term introduced by Sendik and
Cohen-Or \citet{sendik+cohen-or17} derived from the autocorrelation of the
feature maps that was developed to capture periodic structure in image
textures.\footnote{Note that Sendik and Cohen-Or \citet{sendik+cohen-or17}
use a variant of the feature map autocorrelation called the structural
matrix, but we find that the autocorrelation works well and is faster to
compute.}  The autocorrelation of the $k$th feature map is
\begin{equation}
  A_{\tau \mu}^k =
  \mathscr{F}^{-1}_{f} \left[ \mathscr{F}_t [F_{t \mu}^k]
  \, \mathscr{F}_t [F_{t \mu}^k]^* \right],
\end{equation}
where $\mathscr{F}_t$ represents the discrete Fourier transform with respect
to time $t$, $^*$ represents complex conjugation, and $\tau$ represents
the lag.  The autocorrelation loss is the sum of the 
normalized Frobenius norms of the squared differences between the target and
synthesized autocorrelation maps:
\begin{equation}
  \label{eq:autocorrelation_loss}
  \textstyle
  \autocorrelationloss = \frac{\sum_{k, \tau, \mu}
  \left( A_{\tau \mu}^k - \widetilde{A}_{\tau \mu}^k \right)^2}
  {\sum_{k, \tau, \mu} \left( \widetilde{A}_{\tau \mu}^k \right)^2}.
\end{equation}
We generally do not expect to encounter rhythmic structure on timescales
longer than a few seconds, and autocorrelations on extremely short
timescales (under 200~ms) are captured within the receptive fields of
individual networks.  Including very short and long lags in the loss tends
to encourage overfitting without adding any useful rhythmic activity to the
texture (this is particularly true for lags near 0 since the autocorrelation
will always be largest there and will therefore be the largest contributor
to $\autocorrelationloss$).  For this reason we only sum over lags of 200~ms
to 2~s.

\subsubsection{Diversity loss}
\label{subsubsec:diversity_loss}

\begin{figure}
  \centering
  \includegraphics[width=8cm]{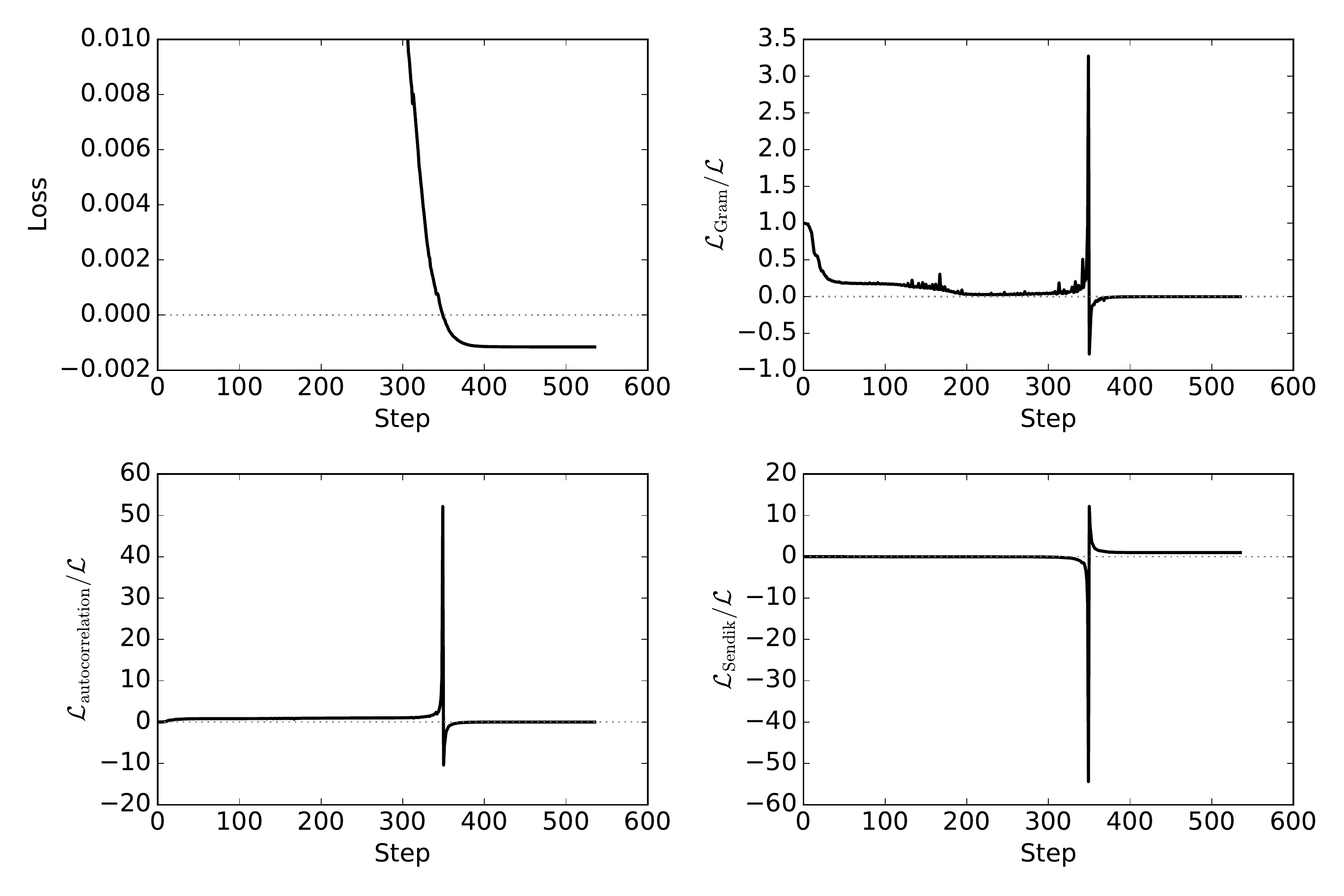}

  \caption{The overall loss and the relative contributions of the three
  components during optimization of a wind chimes texture using
  $\sendikloss$ instead of $\diversityloss$.  This diversity term can lead
  to negative losses, which in turn makes optimization difficult when the
  loss passes through zero.  In part to avoid these instabilities we propose
  a diversity term of the form Eq.~\ref{eq:diversity_term}.}

  \label{fig:diversity_instability}
\end{figure}

\begin{figure}
  \centering
  \includegraphics[width=7cm]{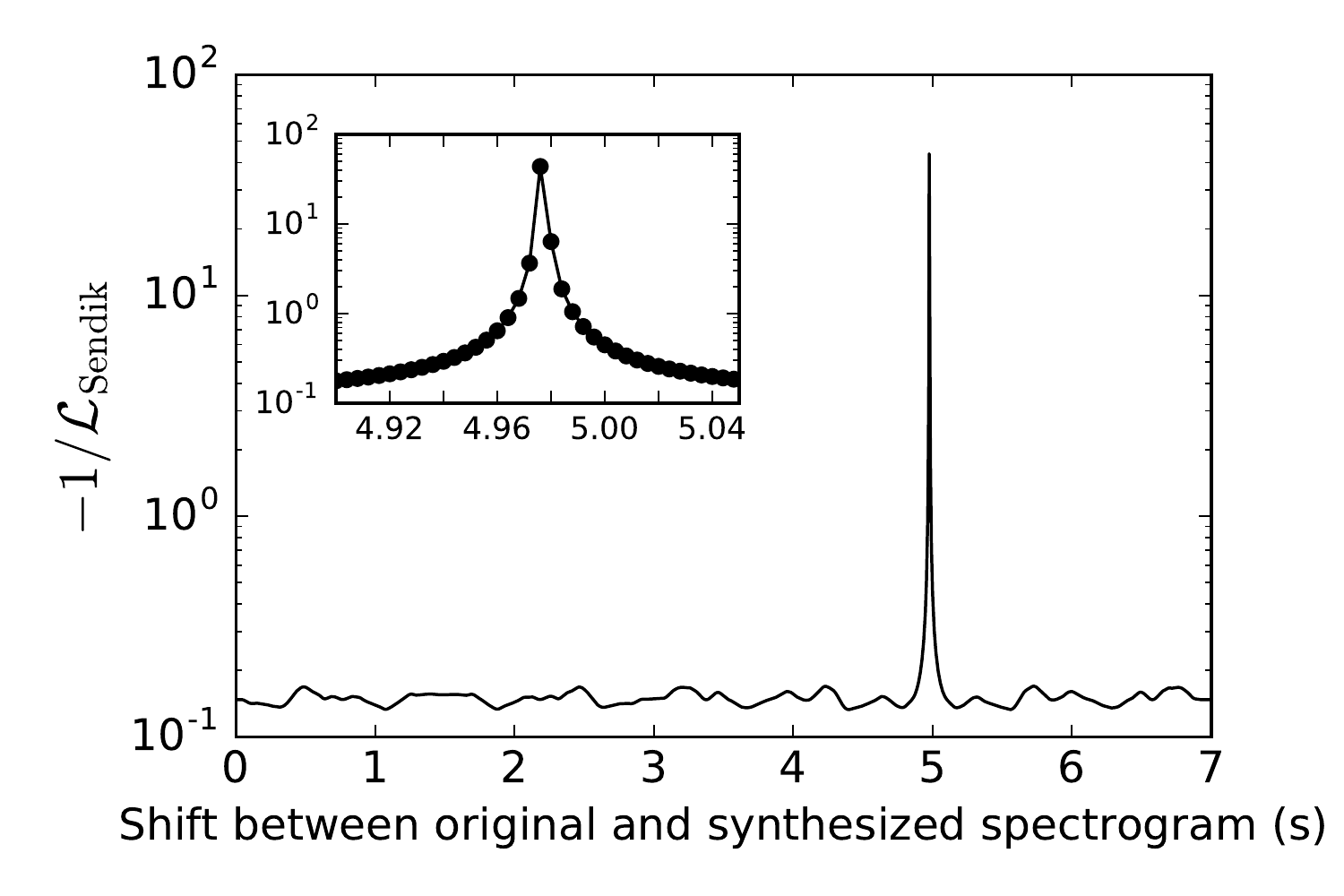}

  \caption{The inverse negative diversity term used by Sendik and Cohen-Or
  \citet{sendik+cohen-or17} as a function of a synthesized texture shifted
  in time.  The synthesized texture closely matches the original texture,
  but is shifted in time by about five seconds.  $\sendikloss$ fails to
  capture this effect as demonstrated by the sharp peak.  Inset zooms in on
  the peak to show that the peak is resolved.}

  \label{fig:shifted_diversity}
\end{figure}

As we show in Sec.~\ref{subsubsec:diversityloss_experiments}, a downside of
using the previous two loss terms alone is that they tend to reproduce the
original texture exactly.  Sendik and Cohen-Or \citet{sendik+cohen-or17}
proposed a diversity term for image texture synthesis of the form
\begin{equation}
  \label{eq:sendik_diversity}
  \textstyle
  \sendikloss = - \sum_{k, t, \mu}
  \left(F_{t \mu}^k - \widetilde{F}_{t \mu}^k \right)^2,
\end{equation}
which is maximized when the two feature maps match exactly.  We found that
this diversity term has two shortcomings: first, because this term can
become arbitrarily negative, it can dominate the total loss and destabilize
the optimization (see Fig.~\ref{fig:diversity_instability}); second, we find
that this loss has a tendency to reproduce the original input, but slightly
shifted in time (see Fig.~\ref{fig:shifted_diversity}).  To address these
two issues, we propose the following shift-invariant diversity term:
\begin{equation}
  \label{eq:diversity_term}
  \textstyle
  \diversityloss = \max_s
  \left( \frac{ \sum_{k, t, \mu} \left( \widetilde{F}_{t \mu}^k \right)^2}
  { \sum_{k, t, \mu} \left(F_{t + s, \mu}^k - \widetilde{F}_{t\mu}^k \right)^2 }
  \right),
\end{equation}
where the shift $s$ can take on values ranging from 0 to $T-1$.  In other
words, we compute the negative inverse of the diversity term of
Eq.~\ref{eq:sendik_diversity} for all possible relative shifts between the
original and synthesized textures and then take the maximum.  Since
computing this loss for all possible shifts is computationally expensive, we
compute this loss in steps of 50 frames, cycling through different sets of
frames in each step of the optimization process, along with computing the
loss for the shifts which yielded the largest loss in the last 10
optimization steps.

\subsection{Optimization}
\label{subsec:optimization}

\begin{figure*}
  \centering
  \includegraphics[width=16cm]{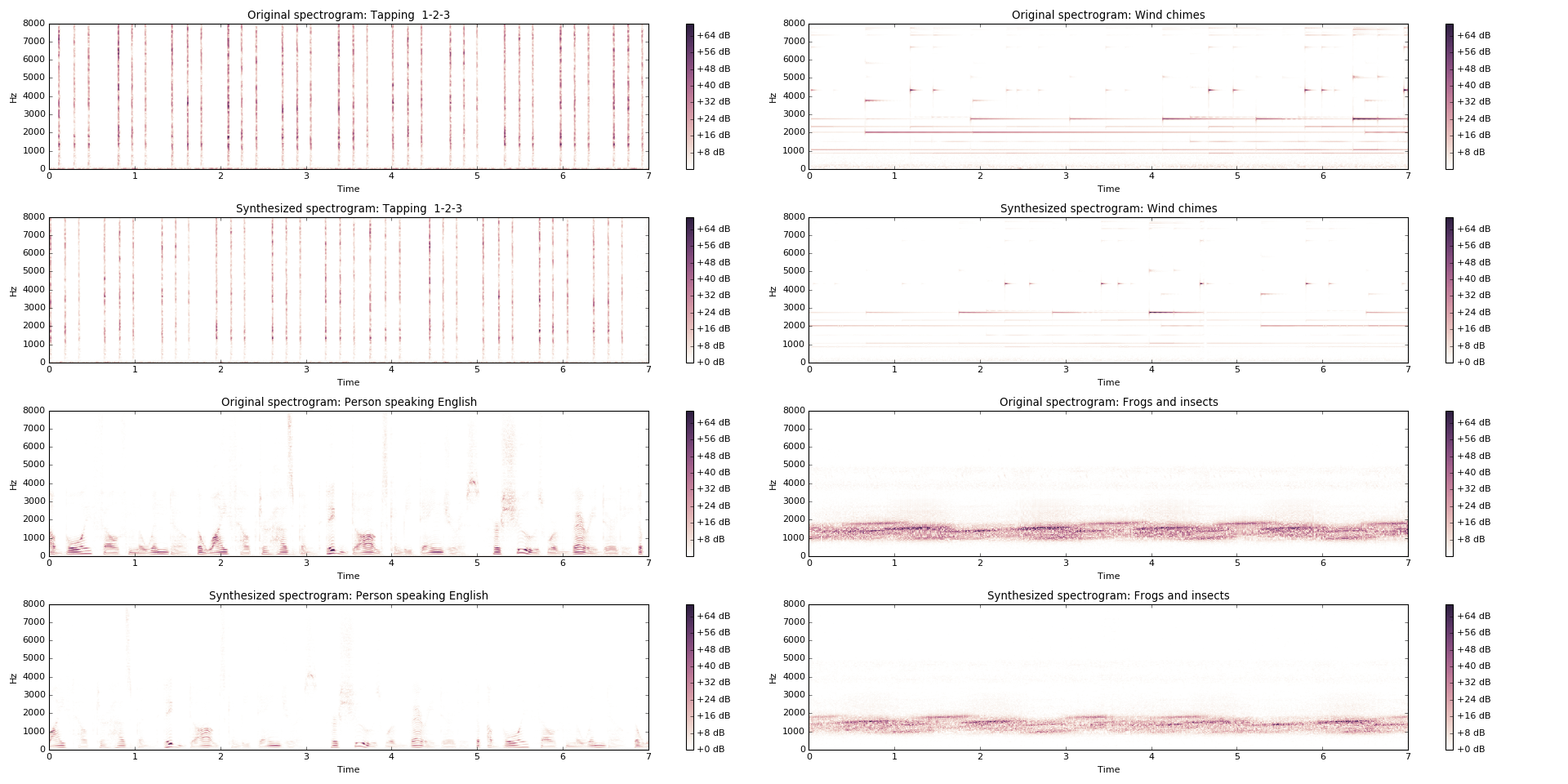}

  \caption{Four pairs of synthesized audio textures with the originals.
  These textures include pitched audio (wind chimes, upper right), rhythmic
  audio (tapping, upper left), speech (lower left), and natural sounds
  (lower right).}

  \label{fig:example_textures}
\end{figure*}

\begin{figure}
  \centering
  \includegraphics[width=8cm]{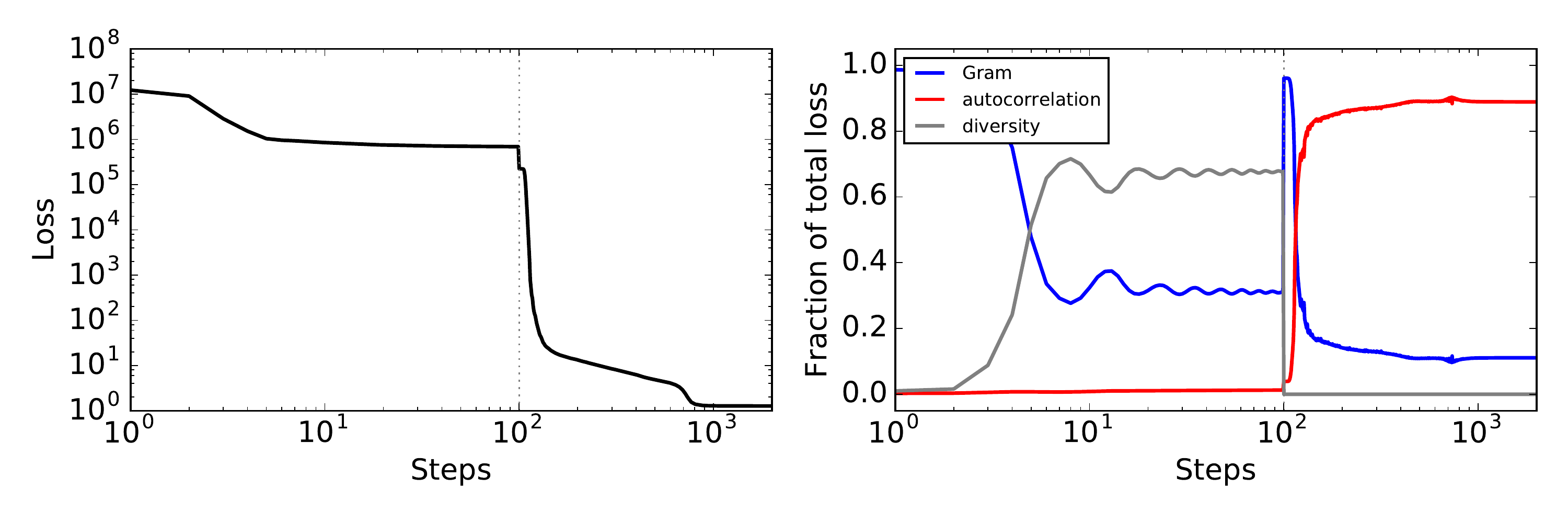}

  \caption{Left panel: loss during optimization of the wind chimes texture.
  Right panel: the fraction of the total loss each of the three terms
  contributes during optimization.}

  \label{fig:training_curves}
\end{figure}

We find that L-BFGS-B \cite{zhu+97} works well to minimize the loss and
obtain high quality audio textures.  We optimized for 2000 iterations and
used 500 iterations of the Griffin-Lim algorithm.  We furthermore found it
useful to include the diversity loss term for only the first 100 iterations;
by this point the optimizer had found a nontrivial local optimum, and
continuing to incorporate the diversity loss reduced texture quality.
Spectrograms of four synthesized textures are shown in
Fig.~\ref{fig:example_textures}.  We show the relative values of the various
loss terms during optimization in Fig.~\ref{fig:training_curves}.
Corresponding audio for all spectrograms, along with supplementary
information, can be found at
\url{https://antognini-google.github.io/audio_textures/}.

\section{Experiments}
\label{sec:experiments}

\subsection{Quantitative evaluation of texture quality}
\label{subsec:vggish_score}

Quantitatively evaluating the quality of generative models is difficult.
Salimans \citet{salimans+16} developed a useful quantitative metric for
comparing generative adversarial networks based on the Kullback-Liebler
divergence of the label predictions given by the Inception classifier
\cite{szegedy+15} between the sampled images and the original dataset.  We
adapt this ``Inception score'' to assess the quality of our audio textures
compared to other methods.  Rather than Inception, we use the ``VGGish''
CNN\footnote{Available from
\url{https://github.com/tensorflow/models/tree/master/research/audioset}.}
that was trained on AudioSet.\footnote{VGGish produces 128 dimensional
embeddings rather than label predictions.  To obtain label predictions we
trained a set of 527 logistic regression classifiers on top of the AudioSet
embeddings (AudioSet's 527 classes are not mutually exclusive.)  We trained
for 100,000 steps with a learning rate of 0.1 and achieved a test accuracy
of 99.42\% and a test cross entropy loss of 0.0584.} The motivation for our
``VGGish score'' is that the label predictions produced by the VGGish model
should match between the original and synthesized textures.  To this end, we
define the score as
\begin{equation}
  \mathcal{S}_{\textrm{{\scriptsize VGGish}}} \equiv
  \exp \left[ \mathbb{E}_x \left[ \textrm{KL} \left(
  p_\mathrm{{\scriptsize VGGish}}(y | \widetilde{x}) \,||\,
  p_\mathrm{{\scriptsize VGGish}}(y | x) \right) \right] \right],
\end{equation}
where $y$ represents the VGGish label predictions and $x$ represents the
texture audio.  We compute $\vggishscore$ over the 168 textures used by
McDermott and Simoncelli \citet{mcdermott+simoncelli11}.  These textures
span a broad range of sound, including natural and artificial sounds,
pitched and non-pitched sounds, and rhythmic and non-rhythmic sounds.  We
compare this VGGish score between our models optimized with different loss
terms and the approaches used by Ulyanov and Lebedev
\citet{ulyanov+lebedev16} and McDermott and Simoncelli
\citet{mcdermott+simoncelli11} in Table~\ref{tbl:scores}, separating out the
scores for pitched and rhythmic textures.  We also compare an
autocorrelation score and a diversity score computed from the generated
spectrograms discussed in
Sections~\ref{subsubsec:autocorrelation_experiments}
and~\ref{subsubsec:diversityloss_experiments}, respectively.

\begin{table*}
\centering

  \caption{A comparison of scores between our model and other work.}

\begin{small}
\begin{tabular}{lrrrrrrrrr}
  \toprule

  & \multicolumn{3}{c}{VGGish $(\times 10^{-4})$}
  & \multicolumn{3}{c}{Autocorrelation}
  & \multicolumn{3}{c}{Diversity} \\

  & Rthm. & Ptch. & Other
  & Rthm. & Ptch. & Other
  & Rthm. & Ptch. & Other \\

  \midrule

  Spectrograms recovered via Griffin-Lim
  & 9.7   & 12.6  & 7.1
  & 7.4   & 0.54  & 2.9
  & 21.4  & 29.7  & 22.7  \\

  McDermott and Simoncelli \citet{mcdermott+simoncelli11}
  & 16.7  & 33.2  & 8.3
  & 542.0 & 408.1 & 421.9
  & \textbf{1.6}  & \textbf{1.6}   & \textbf{2.0} \\

  Ulyanov and Lebedev \citet{ulyanov+lebedev16}
  & 13.4  & 26.8  & 10.0
  & 40.6  & 23.3  & 27.4
  & 2.9   & 3.0   & 3.3 \\

  $\gramloss$
  & \textbf{9.9} & \textbf{16.8}  & \textbf{7.3}
  & 29.0         & 9.7            & 6.5
  & 2.4          & 3.0            & 3.5 \\

  $\gramloss + \autocorrelationloss$
  & 17.8  & 21.3  & 17.9
  & 13.3  & 7.4   & 15.6
  & 3.4   & 5.4   & 5.0  \\

  $\gramloss + \autocorrelationloss + \diversityloss$ $(\beta = 10^{-5})$
  & 14.5  & 23.0          & 12.2
  & 13.0  & \textbf{2.3}  & 7.2
  & 3.8   & 6.8           & 4.4 \\

  $\gramloss + \autocorrelationloss + \diversityloss$ $(\beta = 10^{-3})$
  & 14.9           & 19.0  & 10.0
  & \textbf{4.7}   & 3.7   & \textbf{7.1}
  & 5.0            & 4.9   & 3.9 \\

  \bottomrule
\end{tabular}
\end{small}

\label{tbl:scores}
\end{table*}

The best VGGish scores are obtained by using $\gramloss$ alone.  As
expected, adding $\autocorrelationloss$ substantially reduces the
autocorrelation score, though at the cost of increasing the diversity score,
and adding a larger weight to $\diversityloss$ generally reduces the
diversity score.  The lowest diversity scores are obtained by McDermott and
Simoncelli \citet{mcdermott+simoncelli11}, though at the cost of
substantially higher autocorrelation scores and relatively large VGGish
scores for pitched textures.

It is unsurprising that adding $\diversityloss$ reduces the VGGish score
because introducing any diversity will generally reduce the VGGish score
(the model could achieve a perfect VGGish score simply by copying the
original input).  It is, however, surprising that adding
$\autocorrelationloss$ alone also reduces the VGGish score.  Although
introducing $\autocorrelationloss$ qualitatively seems to increase
overfitting, this overfitting occurs on very long timescales (i.e., the
model will reproduce several seconds that sound very similar to the original
audio).  Introducing $\autocorrelationloss$ seems to make the optimization
process more difficult for timescales much shorter than the minimum lag
considered by $\autocorrelationloss$, which leads to lower quality on short
timescales and thus higher VGGish scores.

\subsection{Effect of the different loss terms}

\subsubsection{Autocorrelation loss}
\label{subsubsec:autocorrelation_experiments}

\begin{figure}
  \centering
  \includegraphics[width=8cm]{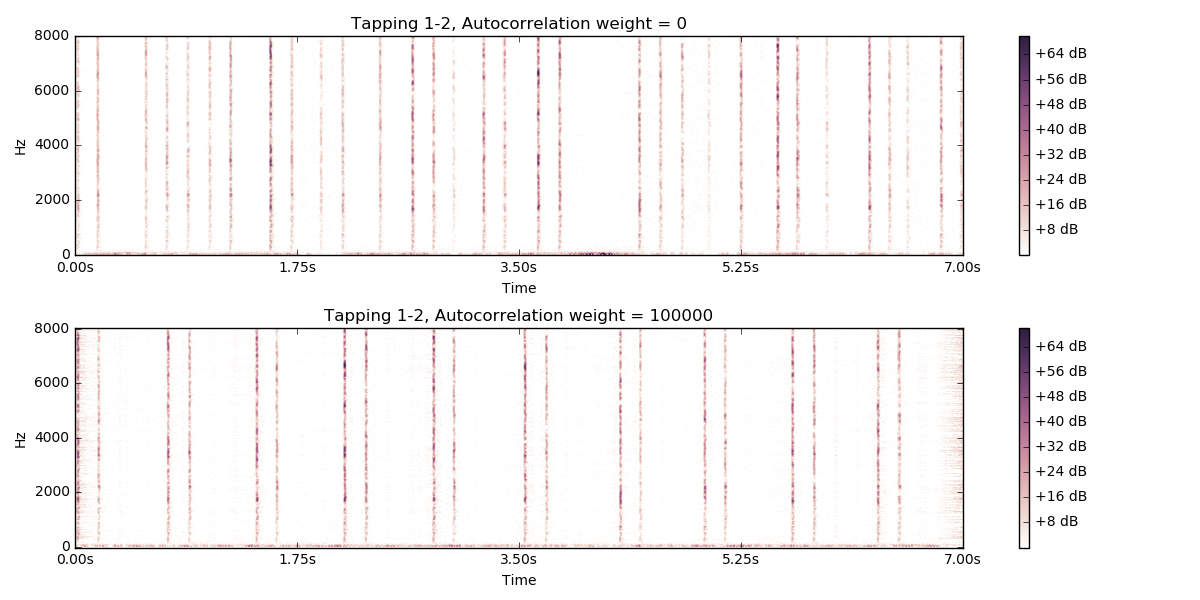}

  \caption{Spectrograms of a rhythmic tapping texture synthesized with
  different weights for $\autocorrelationloss$.  Rhythm is only reproduced
  when the weight on $\autocorrelationloss$ is sufficiently large.}

  \label{fig:autocorrelation_weight}
\end{figure}

\begin{figure*}
  \centering
  \includegraphics[width=16cm]{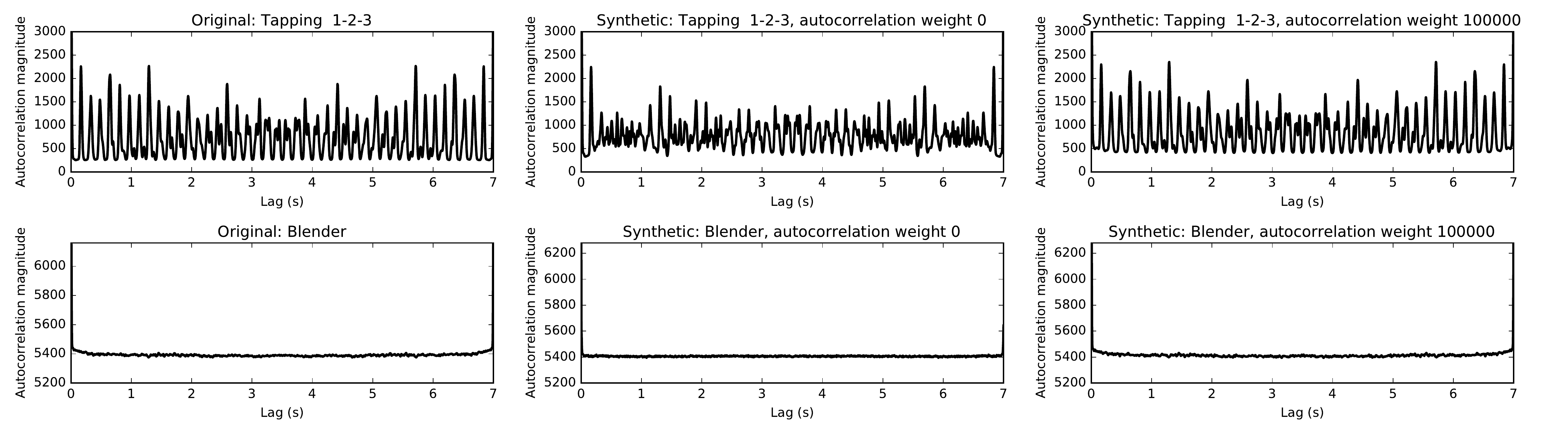}

  \caption{Autocorrelation functions of a rhythmic (top row) and
  non-rhythmic (bottom row) texture for the original (left column) and two
  weights on $\autocorrelationloss$.  Whereas the non-rhythmic texture has a
  flat autocorrelation function, the autocorrelation function of the
  rhythmic texture displays structure that is reproduced only when the
  weight on $\autocorrelationloss$ is large.}

  \label{fig:autocorrelation_matching}
\end{figure*}

To demonstrate the necessity of $\autocorrelationloss$ we synthesize
textures with a variety of values of $\alpha$.  For simplicity we set $\beta
= 0$ in these experiments (i.e., we exclude $\diversityloss$ from the total
loss).  We show spectrograms for a highly rhythmic tapping texture using two
different values of $\alpha$ in Fig.~\ref{fig:autocorrelation_weight}.  If
the weight of $\autocorrelationloss$ is small, the synthesized textures
reproduce tapping sounds which lack the precise rhythm of the original. Only
when $\alpha$ is sufficiently large is the rhythm reproduced.  We further
demonstrate that minimizing $\autocorrelationloss$ reproduces rhythms by
showing in Fig.~\ref{fig:autocorrelation_matching} the autocorrelation
functions of the spectrograms of a rhythmic and non-rhythmic
texture.\footnote{Note that this is \emph{not} directly minimized by
minimizing Eq.~\ref{eq:autocorrelation_loss} which is a function of the CNN
features, not the spectrogram itself.} We furthermore compute the squared
loss between the autocorrelation of each synthesized texture and its target
texture, normalized to the Frobenius norm of the autocorrelation of the
target texture.  We present these scores in Table~\ref{tbl:scores}.
Qualitatively we find that, as expected, $\autocorrelationloss$ is most
important in textures with substantial rhythmic activity and so it is useful
to use a relatively large value for $\alpha$ for these rhythms.  For textures
without substantial rhythmic activity we find that a smaller choice of
$\alpha$ produces higher quality textures.

\subsubsection{Diversity loss}
\label{subsubsec:diversityloss_experiments}

\begin{figure*}
  \centering
  \includegraphics[width=16cm]{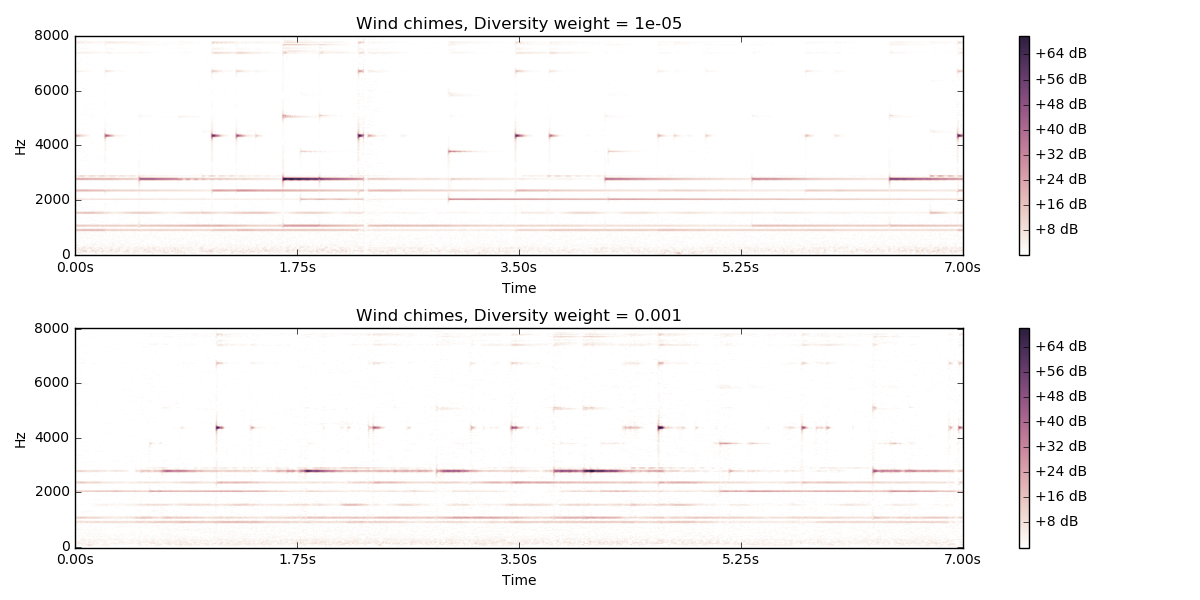}

  \caption{Spectrograms synthesized with different weights $\beta$ on
  $\diversityloss$ for two non-stationary sounds: wind chimes (left)
  and speech (right).}

  \label{fig:diversity_weight}
\end{figure*}

\begin{figure}
  \centering
  \includegraphics[width=7cm]{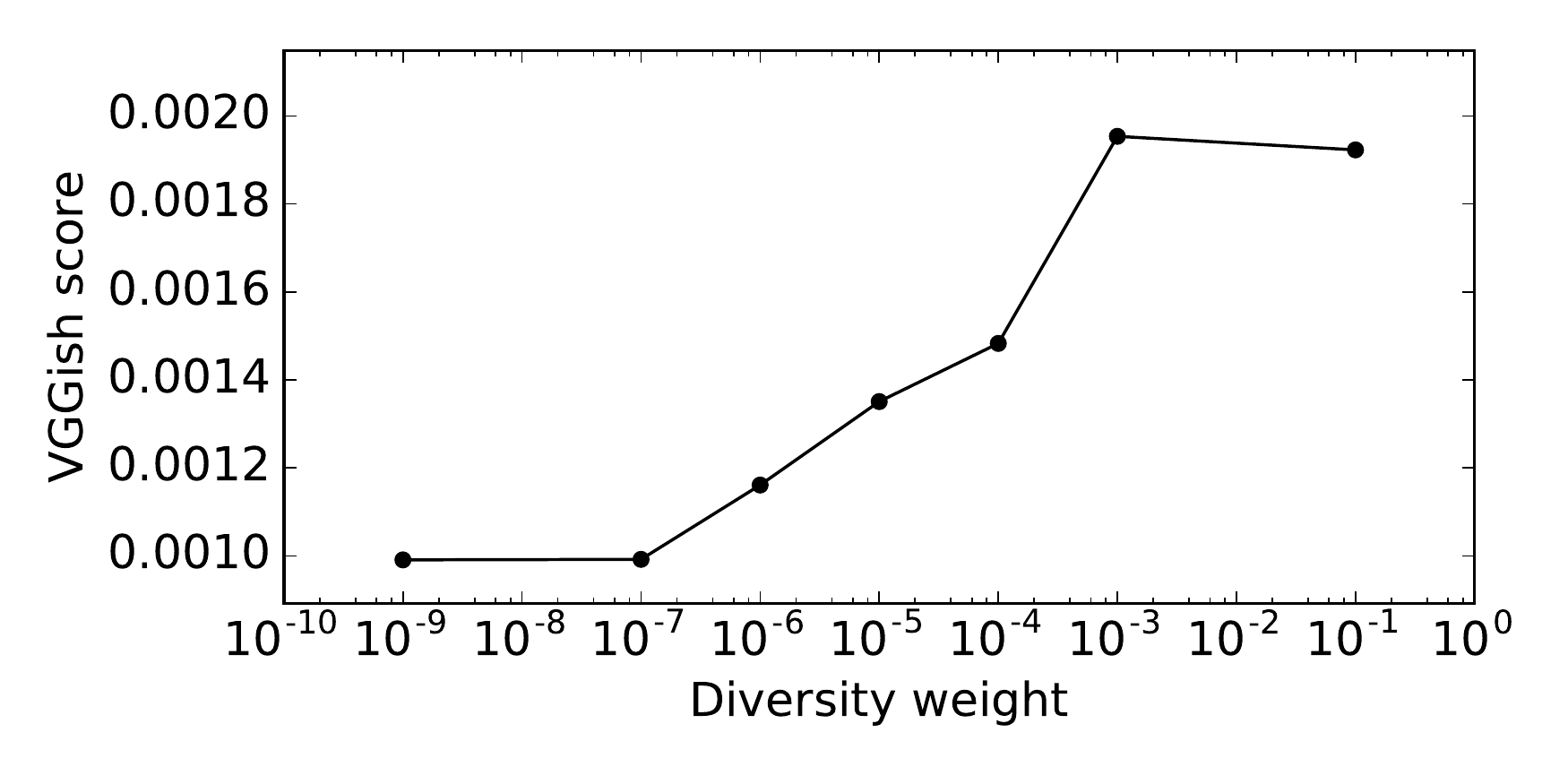}

  \caption{The diversity-quality trade-off in texture synthesis.  The VGGish
  score is a rough proxy for texture quality, with lower scores representing
  higher quality textures.  As the diversity weight increases, the
  average quality of the textures decreases.}

  \label{fig:diversity_quality}
\end{figure}

To demonstrate the effect of $\diversityloss$ we synthesize textures with a
variety of values of $\beta$, keeping $\alpha$ fixed to $10^3$.  We show in
Fig.~\ref{fig:diversity_weight} spectrograms synthesized with two different
values of $\beta$ for two highly structured textures: wind chimes and
speech.  Smaller values of $\beta$ generally reproduce the original texture
but shifted in time (about 2~s for the wind chimes and about 3.25~s for
speech).  Larger values of $\beta$ produce spectrograms which are not simple
translations of the original input, but the quality of the resulting audio
is much lower.  In the case of the wind chimes the chimes do not have the
hard onset in the original, and in the case of speech the voice is echoey
and superimposes different phonemes.  This is an instance of a more general
diversity-quality trade-off in texture synthesis.  In
Fig.~\ref{fig:diversity_quality} we show the VGGish score (a rough proxy for
texture quality) vs.~the weight on the diversity term.  As the weight on the
diversity term increases, the average quality decreases.
We furthermore calculate the diversity loss on the spectrograms themselves to
get a diversity score and present these scores in Table~\ref{tbl:scores}.

We find that it is crucial to tune the loss weights for different texture
classes in order to obtain the highest quality textures.  Large $\alpha$ and
large $\beta$, for example, is especially important for reliably generating
rhythmic textures.  Pitched audio generally requries a smaller choice of
$\alpha$ and $\beta$.  For non-textured audio like speech and music, high
quality audio is only obtained with a large $\alpha$ and small $\beta$,
which will only reproduce the original with some shift; since these kinds of
audio do not obey the assumptions set out in
Section~\ref{sec:preliminaries}, any set of weights that does not reproduce
the original will produce low quality audio.

\subsection{Neural network architecture}
\label{subsec:nn_architecture}

\begin{figure*}
  \centering
  \includegraphics[width=16cm]{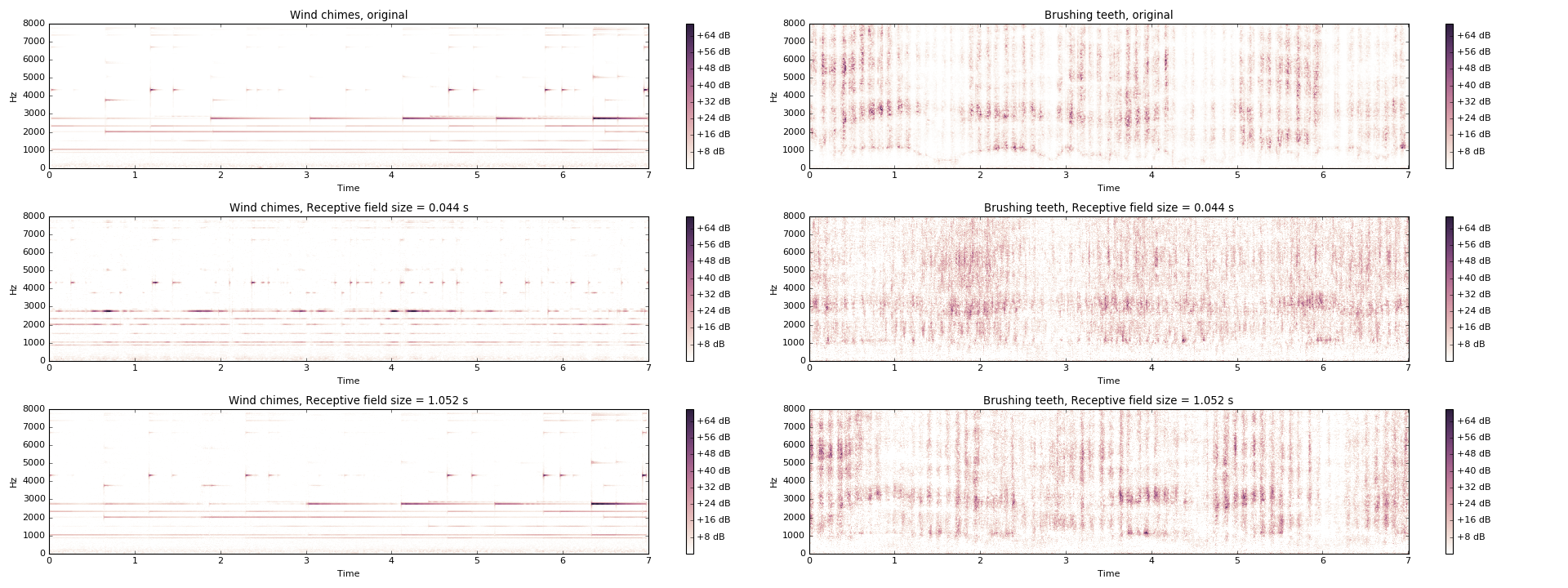}

  \caption{Spectrograms synthesized from convolutional networks with different
  maximum receptive field sizes.}

  \label{fig:conv_width}
\end{figure*}

The receptive field size of the convolutional kernel has a strong effect on
the quality and diversity of the synthesized textures.  We show in
Fig.~\ref{fig:conv_width} the effect of changing the receptive field size
for two textures.  To do this, we use the same set of single layer CNNs with
exponentially increasing kernel sizes, but varying the maximum kernel size
from 2 frames to 8.  CNNs with very small receptive fields produce novel,
but poor-quality textures that fail to capture long-range structure.
Networks with large receptive fields tend to reproduce the original.  This
is an example of the quality-diversity trade-off in texture synthesis.

\begin{figure*}
  \centering
  \includegraphics[width=16cm]{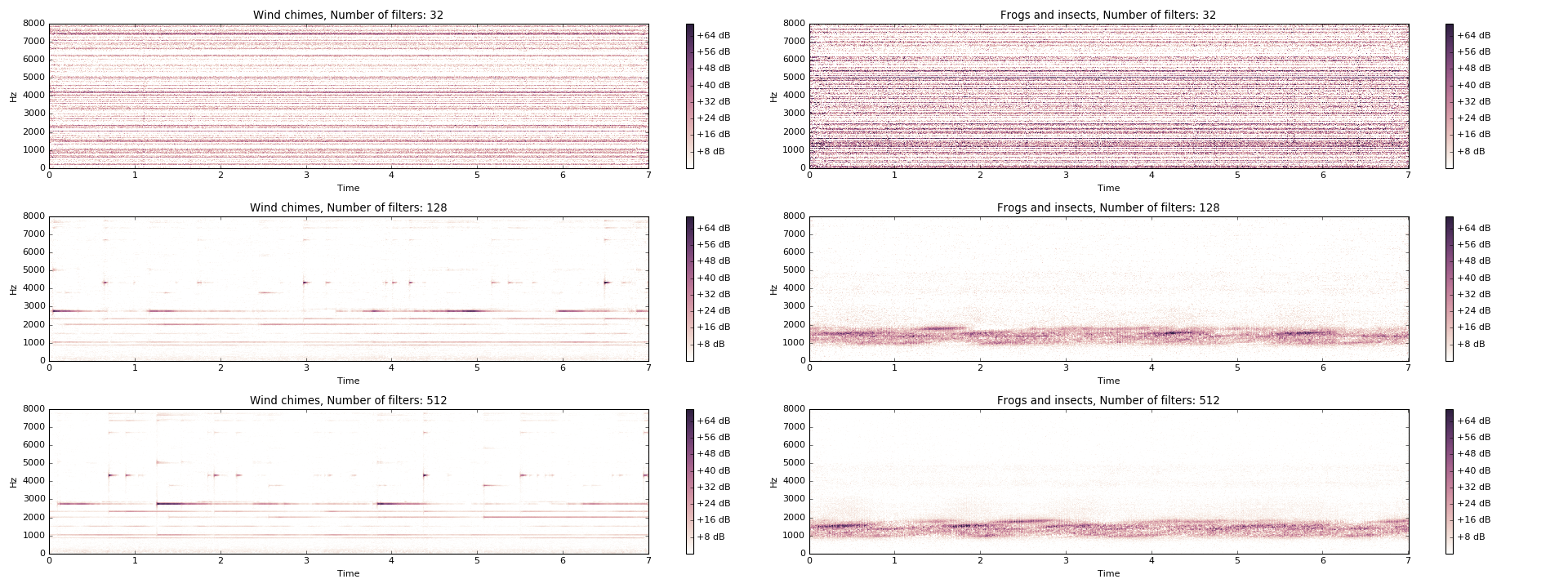}

  \caption{Spectrograms for two textures synthesized with varying numbers of
  filters in the convolutional layers.  With 32 filters, the textures have
  isolated power in some frequencies and not others, but have not developed any
  temporal structure.  It is not until there are 128 filters that the textures
  introduce sounds that vary over time.  However, for complex textures like wind
  chimes, only 512 filters are sufficient to produce the hard onsets of bells.
  See Fig.~\ref{fig:example_textures} for the originals.}

  \label{fig:n_filters}
\end{figure*}

Another design choice we consider is the number of filters in the each
network.  We show in Fig.~\ref{fig:n_filters} the results of using 32, 128,
and 512 filters to synthesize two textures.  Note that because there are six
CNNs in all with varying kernel sizes, the total number of activations
varies from 192 to 3072.  At least 128 filters are necessary to get
reasonable textures, but the quality continues to improve with 512 filters.

\begin{figure*}
  \centering
  \includegraphics[width=16cm]{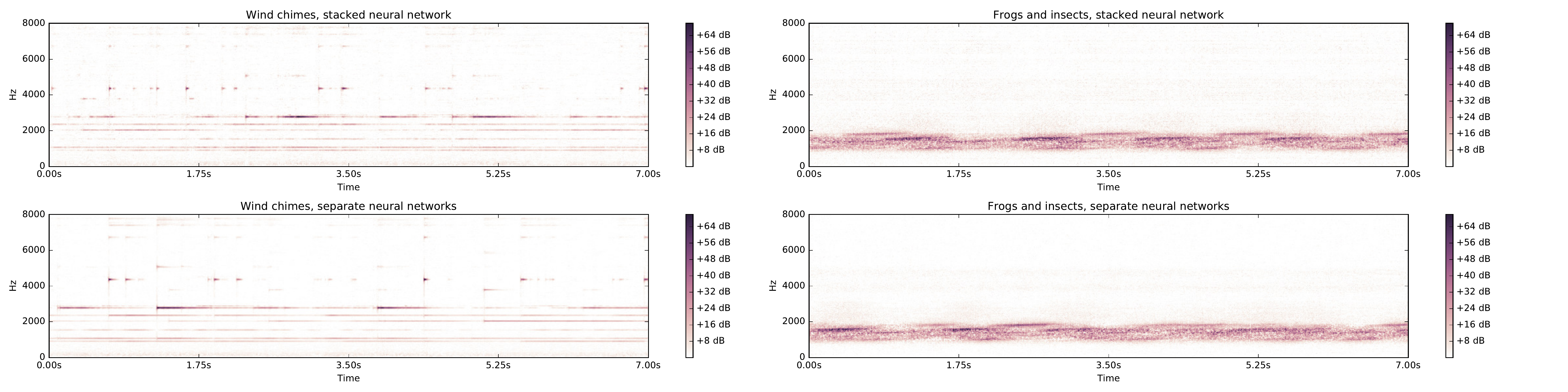}

  \caption{Spectrograms for two textures synthesized with a neural network
  that consists of six convolutional layers stacked on top of each other,
  each with a receptive field size of 2 frames and separated by average
  pooling layers with a pooling size of 2 and a stride of 2.  We compare
  these spectrograms to spectrograms synthesized with the six separate
  neural networks of varying receptive field sizes used elsewhere in this
  paper.  The distribution of receptive field sizes in both neural networks
  is the same.  See Fig.~\ref{fig:example_textures} for the originals.}

  \label{fig:stacking}
\end{figure*}

We considered stacking six convolutional layers on top of each other, each
with a receptive field of 2 and separated by an average pooling layer with a
pool size of 2 and a stride of 2.  This network has the same distribution of
receptive field sizes as the six separate networks, but the input to each
layer here must pass through the (random) filters of all the earlier layers.
We compare spectrograms generated with this network to the six separate
networks that we use elsewhere in
Fig.~\ref{fig:stacking}.  The only effect of stacking the layers is a
modest degradation in the quality of long-range sounds, best seen in
the wind chimes texture.

\section{Discussion}
\label{sec:discussion}

\subsection{Towards compelling audio style transfer}
\label{subsec:style_transfer}

After Gatys et al. \citet{gatys+15a} developed the Gram-based approach for
texture synthesis, a natural extension was to propose a similar technique
for artistic style transfer \cite{gatys+15b} in which there is an additional
content term in the loss which is minimized by matching the high-level
features of a second image.  There have now been a variety of impressive
results for image style transfer \cite{chen+16, dumoulin+16, ulyanov+16}.
There have been a few attempts to extend these techniques to audio style
transfer \cite{ulyanov+lebedev16, grinstein+17, barry+kim18}, and while the
results are plausible, they are underwhelming compared to the results in the
image domain.  What makes the audio domain so much more challenging?

The first issue is that it is unclear what is meant by ``style transfer.''
The simplest form of style transfer would be to take a melody played on one
instrument and make it sound as though it were played on another; or
similarly voice conversion, i.e., taking audio spoken by one person and
making it sound as though it had been spoken by another \cite{chorowski+17}.
While this simpler version of style transfer is still an open problem, the
more interesting and far more difficult form of style transfer would be
taking a melody from one genre (e.g., a Mozart aria) and transforming it
into one from another (e.g., a jazz song) by keeping the broad lyric and
melodic structure, but replacing the instrumentation, ornamentation,
rhythmic patterns, etc., characteristic of one genre with those of another.
It is plausible that the simpler form of style transfer can be accomplished
with careful design choices in the convolutional architecture
\cite{chorowski+17}.  But it does not appear that such an approach can work
for the more complicated kind of style transfer.

To understand why, it is worth comparing the features learned by deep CNNs
trained on image vs.~audio data.  In the image domain there is a well
defined feature hierarchy, with lower layers learning simple visual patterns
like lines and corners, and later layers learning progressively more
complicated and abstract features like dog faces and automobiles in the
final layers \cite{olah+17}.  By contrast, CNNs in the audio domain have
been far less studied.  Dieleman \citet{dieleman14} analyzed patterns in the
feature activations of a deep CNN trained on one million songs from the
Spotify corpus in van den Oord \citet{van-den-oord+13}.  Dieleman
\citet{dieleman14} found that features in lower layers identified local
stylistic and melodic features, e.g., vibrato singing, vocal thirds, and
bass drum.  Features in later layers identified specific genres, e.g.,
Christian rock and Chinese pop.  Whereas in the image domain the feature
activations of the higher layers represent the content of the image (e.g.,
there is a dog in the lower left corner), in the audio domain the later
layers instead represent the overall style.  These differences reflect the
way that these CNNs are trained.  In the image domain, CNNs are explicitly
trained to identify the content of the image.  In the audio domain the CNN
is instead trained to identify the overall style.  This poses a challenge
for style transfer because the ``content'' of the audio consists of the
melody and lyrics, but the CNN is never trained to identify the content and
so it either gets mixed in with other low-level textural features or is not
propagated to later layers at all.  Successful audio style transfer will
require a network that can separate the melodic content of audio from its
stylistic content the way that image classification CNNs can.  The path
forward may instead lie with neural networks trained on transcription or
Query by Singing/Humming tasks.

\section{Conclusions}
\label{sec:conclusions}

We have demonstrated that the approach to texture synthesis described by
Gatys et al.~\citet{gatys+15a} of matching Gram matrices from convolutional
networks can be extended to the problem of synthesizing audio textures.
There are, however, certain differences in the audio domain vs.~the image
domain that require the addition of two more loss terms to produce diverse,
robust audio textures: an autocorrelation term to preserve rhythm, and a
diversity term to encourage the synthesized textures to not exactly
reproduce the original texture.  We test our technique across several
classes of textures like rhythmic and pitched audio and find that tuning the
weights on the autocorrelation and diversity terms is crucial to obtaining
the highest quality textures for different classes.  The choice of
architecture is also important to obtain high quality textures; an ensemble
random convolutional neural networks with a wide range of receptive field
sizes allows the model to capture features that occur on timescales across
many orders of magnitude.  Finally, we show that this method has a trade-off
between the diversity and the quality of the results.

\section*{Acknowledgments}

The authors are grateful to Josh McDermott for providing the synthesized
textures using the technique of McDermott and Simoncelli
\citet{mcdermott+simoncelli11} for comparison with the technique in this
paper.  The authors thank Rif A.~Saurous for helpful comments on the
manuscript.

\bibliography{refs}
\bibliographystyle{IEEEtran}

\vspace{-0.5in}
\begin{IEEEbiography}[{\includegraphics[width=1in,height=1.25in,clip,keepaspectratio]{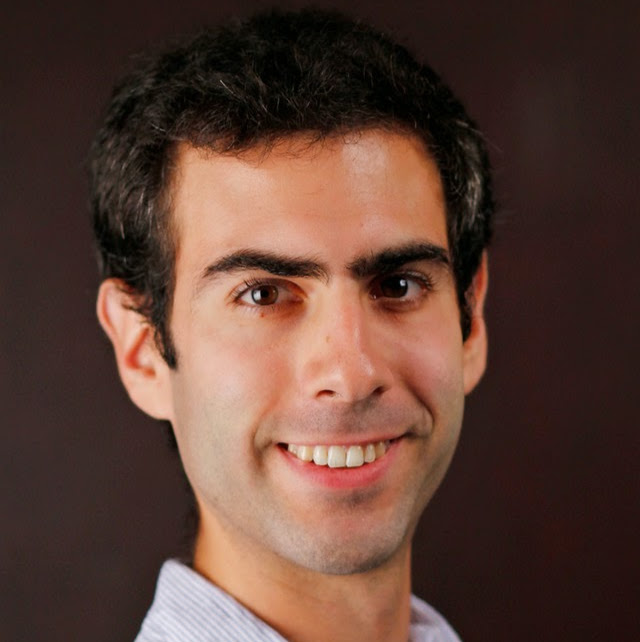}}]
  {Joseph M.~Antognini} is a Google AI resident.  At Google he has worked on
  applying machine learning to audio synthesis and studying the foundations
  of machine learning.  He received his Ph.D.~in astronomy from The Ohio
  State University in 2016.
\end{IEEEbiography}
\vspace{-0.6in}
\begin{IEEEbiography}[{\includegraphics[width=1in,height=1.25in,clip,keepaspectratio]{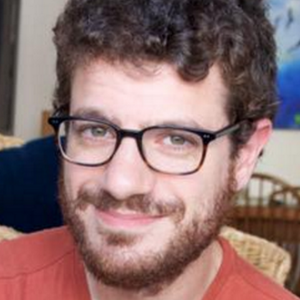}}]
  {Matthew D.~Hoffman} is a senior research scientist at Google. His main
  research focus is in probabilistic modeling and approximate inference
  algorithms. He has worked on various applications including music
  information retrieval, speech enhancement, topic modeling, learning to
  rank, computer vision, user interfaces, user behavior modeling, social
  network analysis, digital imaging, and astronomy. He is a co-creator of
  the widely used statistical modeling package Stan.
\end{IEEEbiography}
\vspace{-0.6in}
\begin{IEEEbiography}[{\includegraphics[width=1in,height=1.25in,clip,keepaspectratio]{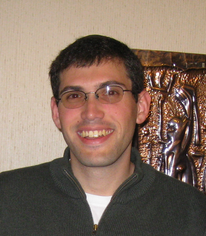}}]
  {Ron J.~Weiss} is a software engineer at Google where he has worked on
  content-based audio analysis, recommender systems for music, and noise
  robust speech recognition. Ron completed his Ph.D. in electrical
  engineering from Columbia University in 2009 where he worked in the
  Laboratory for the Recognition of Speech and Audio. From 2009 to 2010
  he was a postdoctoral researcher in the Music and Audio Research
  Laboratory at New York University.
\end{IEEEbiography}

\end{document}